\documentclass{webofc}
\usepackage[varg]{txfonts}   
\usepackage{hyperref}
\usepackage{amssymb}
\usepackage{amsmath,mathtools}
\usepackage{amsfonts}
\usepackage{latexsym}
\usepackage{multirow}
\usepackage{fixmath}
\usepackage{color}
\usepackage{stackrel}
\usepackage{txfonts}
\usepackage{centernot}
\usepackage{mathabx}
\newcommand{\ols}[1]{\mskip.5\thinmuskip\overline{\mskip-.5\thinmuskip {#1} \mskip-.5\thinmuskip}\mskip.5\thinmuskip} 
\begin{document}

\vspace*{-0.5cm}
\hfill DESY-22-199
\vspace*{0.25cm}

  \title{Dynamical solution of the strong CP problem within QCD ?}

\author{\firstname{\rm Gerrit} \lastname{Schierholz}\inst{1}\fnsep\thanks{\email{gerrit.schierholz@desy.de}}}

\institute{Deutsches Elektronen-Synchrotron DESY\\Notkestr. 85, 22607 Hamburg, Germany}

\abstract{The strong CP problem is inseparably connected with the topology of gauge fields and the mechanism of color confinement, which requires nonperturbative tools to solve it. In this talk I present results of a recent lattice investigation of QCD with the $\theta$ term in collaboration with Yoshifumi Nakamura~\cite{Nakamura:2019ind,Nakamura:2021meh}. The tool we are using to address the nonperturbative properties of the theory is the gradient flow, which is a particular realization of momentum space RG transformations. The novel result is that within QCD the vacuum angle $\theta$ is renormalized, together with the strong coupling constant, and flows to $\theta = 0$ in the infrared limit. This means that CP is conserved by the strong interactions.}
\maketitle
\section{Introduction}
\label{intro}

One of the most intriguing unsolved problems in particle physics is the strong CP problem. While the CP violation observed in $K$ and $B$
meson decays can be accounted for by the phase of the CKM matrix, the baryon asymmetry of the universe cannot be described by this phase alone, suggesting that there are additional sources of CP violation awaiting discovery. QCD allows for a CP-violating term $S_\theta$, called the $\theta$ term, in the action,
\begin{equation}
  S = S_0 + S_\theta \,.
\end{equation}
In Euclidean space-time it reads
\begin{equation}
S_\theta = i\, \theta\, Q\,, \quad Q = \frac{1}{32\pi^2}\, \epsilon_{\mu\nu\rho\sigma} \!\int \! d^4x\; \textrm{Tr}\left[F_{\mu\nu} F_{\rho\sigma}\right]\, \in\, \mathbb{Z}\,,
\label{charge}
\end{equation}
where $Q$ is the topological charge, and $\theta$ is the {\em bare} vacuum angle. Thus, there is the possibility of strong CP violation arising from a nonvanishing value of $\theta$. A finite value of $\theta$ would result in an electric dipole moment $d_n$ of the neutron. To date the most sensitive measurements of $d_n$ are compatible with zero. The current upper bound is $|d_n| < 1.8 \times 10^{-13} e\,\textrm{fm}$~\cite{Abel:2020gbr}, indicating that $\theta$ is anomalously small. Why should a parameter not forbidden by symmetry be essentially zero? This puzzle is referred to as the strong CP problem. 

The electric dipole moment $d_n$ is a measure of (permanent) separation of positive and negative charge in the neutron. According to the upper bound on $d_n$, the separation would have to be less than $10^{-13}$ fm. It would be rather na\"ive to believe that any structure at that scale can be attributed to QCD, and to the topological properties of the theory in particular. On the contrary, the $\theta$ term is expected to have a positive lasting effect at length scales 1/\raisebox{2pt}{$\sqrt[4]{\chi_t}$} $\simeq 1\,\textrm{fm}$, where $\chi_t = \langle Q^2\rangle/V$ is the topological susceptibility, and the typical size of an instanton is $1/3\, \textrm{fm}$~\cite{Shuryak:1995pv}.

A popular view is that the solution of the strong CP problem lies beyond QCD and the Standard Model. Indeed, the first instinct in such a situation is to propose a new symmetry that suppresses CP-violating terms in the strong interactions. Peccei and Quinn~\cite{Peccei:1977hh} concocted such a symmetry in 1977, at the expense of introducing a hitherto undetected particle, the axion. To date I count O(300,000) papers on axions in INSPIRE-HEP. That is all the more reason to investigate the nonperturbative properties of QCD with the $\theta$ term and look for a solution within QCD. This is a task for the lattice. There is even the possibility that the Peccei-Quinn model is inconsistent with color confinement~\cite{GS}. A solution within QCD would mean that QCD does not exist as a viable physical theory unless $\theta = 0 \, [\textrm{mod}\,2\pi]$, or that QCD has an infrared (IR) fixed point at which the vacuum angle renormalizes to $\theta = 0$, or both. As a guideline it is helpful to have some model understanding how the QCD vacuum reacts to the $\theta$ term. Several ideas have been communicated in the literature. An incomplete list is given in~\cite{Coleman:1976uz,tHooft:1981bkw,Cardy:1981qy,Ezawa:1982bf,Knizhnik:1984kn,Wu:1984bi,Samuel:1991cm,Cohen:2018cyj}.

\begin{figure}[h!]
  \vspace*{0.25cm}
  \begin{center}
 \includegraphics[width=6cm]{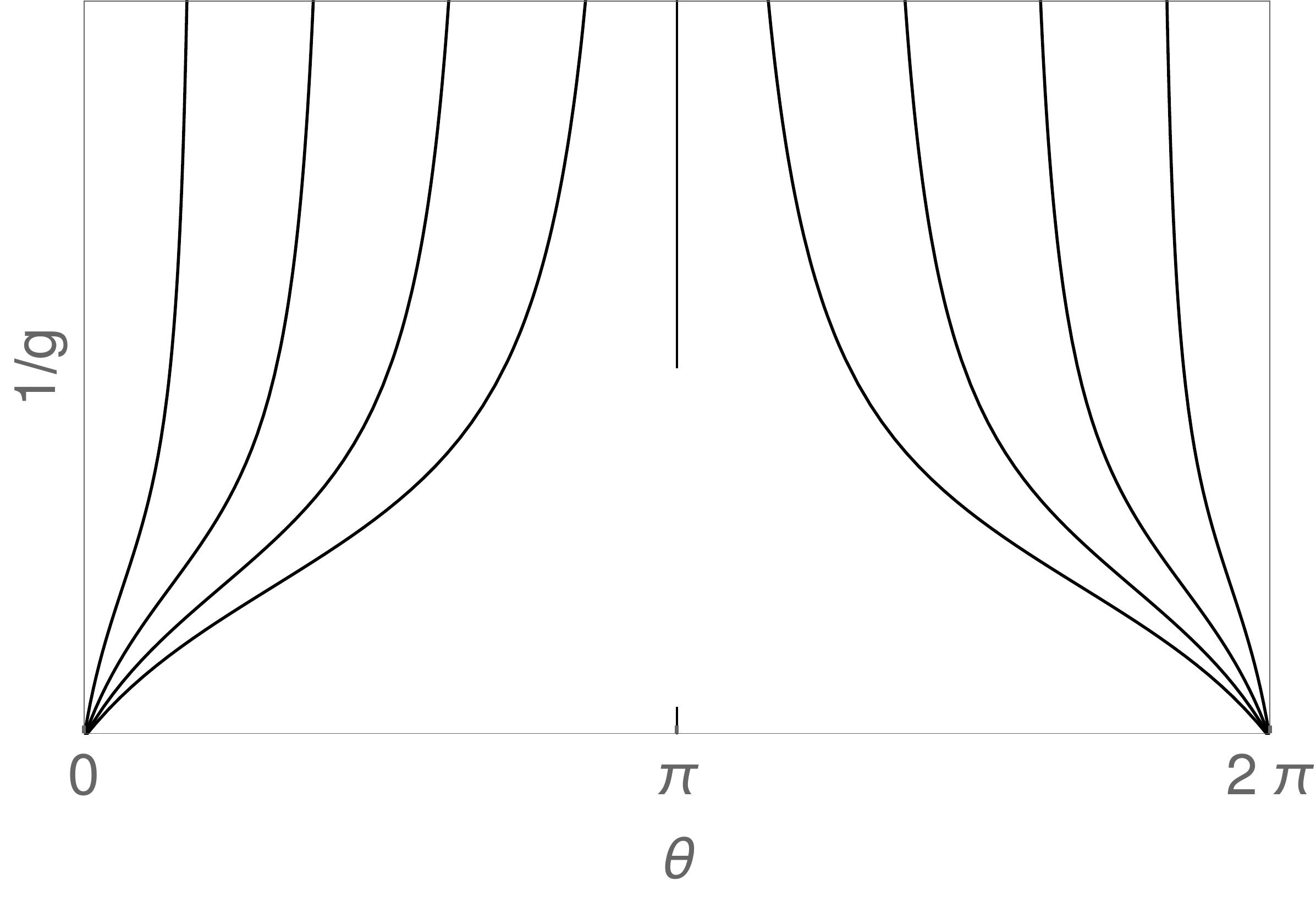}
  \end{center}
  \vspace*{-0.5cm}
\caption{The renormalizaqtion group flow of the $CP^{N-1}$ model at large $N$ for different initial values of $\theta$. The distance scale ($\sim 1/T$) increases from top to bottom.}
\label{fig1}
\vspace*{-0.25cm}
\end{figure}

One source of information are field theories that share the main characteristics of QCD, but lend themselves to (semi-)analytic investigations. A prominent example is the $CP^{N-1}$ model in two dimensions at large $N$, which reflects the fundamental features of the quantum Hall effect~\cite{Pruisken:2000my}. The Hall conductivity $\sigma_{xy}$ has a precise parallel in the vacuum angle $\theta$, $\theta/2\pi \sim \sigma_{xy}$, while the linear conductivity is represented by the inverse coupling, $N/g \sim \sigma_{xx}$. In Fig.~\ref{fig1} I sketch the renormalization group (RG) flow of $1/g$ and $\theta$. The figure shows that any initial value of $\theta$ flows to $\theta=0$ at macroscopic scales. For long, but widely ignored, the quantum Hall effect has served as a model for the solution of the strong CP problem~\cite{Levine:1983vg}. The flow to quantization, $\theta = 0$, appears to be a generic feature of the instanton vacuum as well~\cite{Knizhnik:1984kn}. 

\begin{figure}[t!]
\vspace*{-4cm}
\begin{center}
\hspace*{-0.5cm}\includegraphics[width=.7\linewidth]{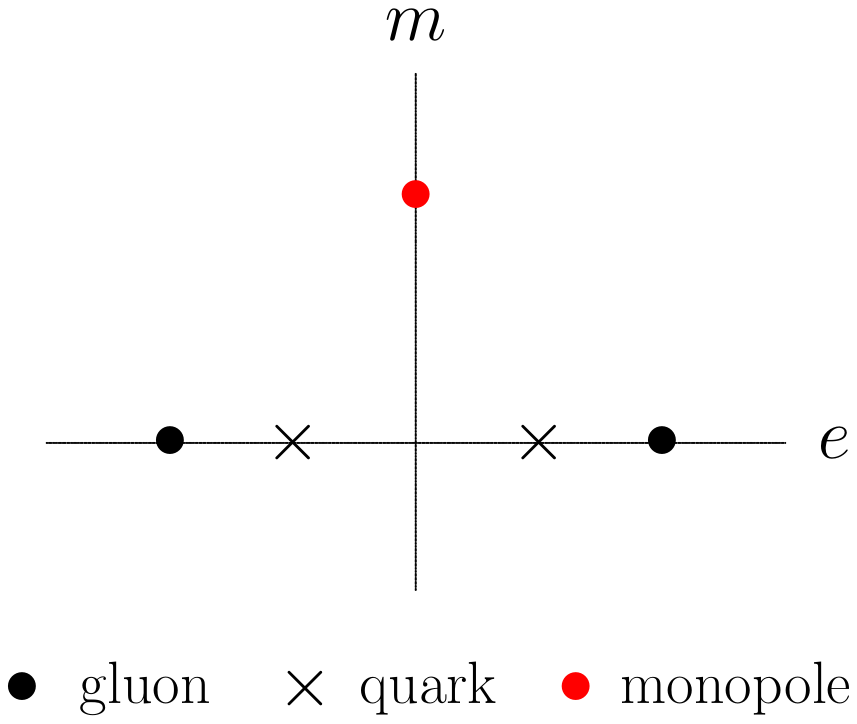}\hspace*{-4.5cm}
  \includegraphics[width=.7\linewidth]{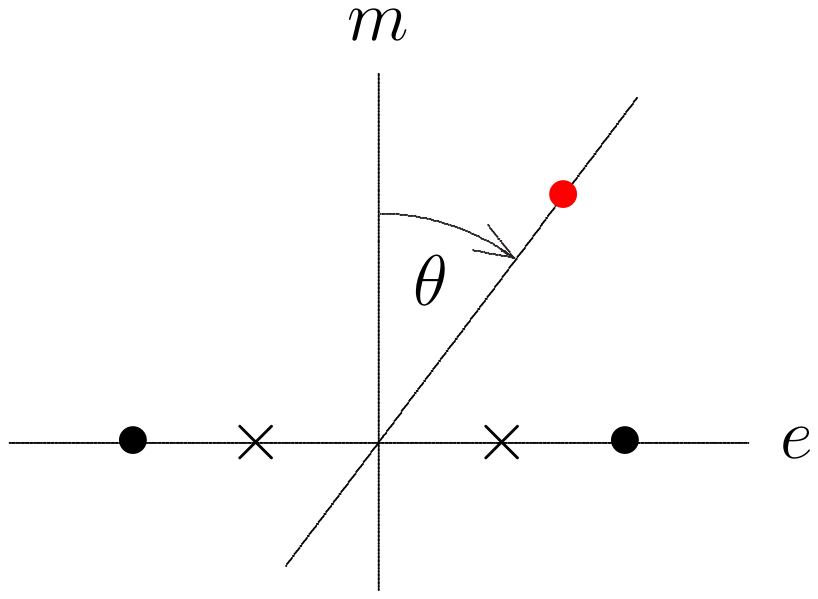}\\
\end{center}
\vspace*{-5cm}
\caption{The color-electric -- color-magnetic charge lattice for vacuum angle $\theta=0$ and $\theta>0$, with regard to the gauge group U(1). Gluons have color-electric charge $\pm 1$, quarks have charge $\pm 1/2$, and monopoles have color-magnetic charge $\pm 1$ in Dirac units.}
\label{fig2}
\vspace*{-0.25cm}
\end{figure}

In QCD we have a fairly good understanding of the intermediate- to long-distance features of the vacuum. A crucial step in this process was to isolate the relevant dynamical variables at the critical distance scale. In a groundbreaking paper 't~Hooft~\cite{tHooft:1981bkw} has argued that the degrees of freedom responsible for confinement are color-magnetic monopoles. This is unfolded by fixing the gauge. An appropriate gauge is the maximal abelian gauge~\cite{Kronfeld:1987vd}, which is an incomplete gauge that leaves the Cartan subgroup $\text{U(1)}\times\text{U(1)} \subset \text{SU(3)}$ unbroken. Monopoles arise from singularities of the gauge condition. Quarks and gluons have color-electric charges with respect to the U(1) subgroups. Confinement occurs when the monopoles condense in the vacuum, by analogy to superconductivity. This has first been verified on the lattice by Kronfeld \textit{et al.}~\cite{Kronfeld:1987ri}, and subsequently tested in countless papers beginning with~\cite{Suzuki:1989gp}. For $\theta$ different from zero the monopoles acquire a color-electric charge $q=\theta/2\pi$~\cite{Witten:1979ey}. Due to the joint presence of gluons and monopoles a rich phase structure is expected to emerge as a function of $\theta$~\cite{tHooft:1981bkw,Cardy:1981qy,Ezawa:1982bf}. In Fig.~\ref{fig2} I show the charge lattice of quarks, gluons and monopoles for $\theta=0$ and $\theta >0$. In the latter case it is expected that the color fields of quarks and gluons are screened by forming bound states with the monopoles. The Debye screening length of a particle of charge $q$ immersed in the conducting vacuum is given by $\lambda_D = \sqrt{E_F/\rho q^2}$, where $E_F$ is the Fermi energy. Here $\rho$ is the monopole density~\cite{DIK:2003alb,Hasegawa:2018qla}. Assuming that $\rho$ does not change significantly for small values of $\theta$, this leads us to
\begin{equation}
  \lambda_D = \sqrt{\frac{E_F}{\rho}}\; \frac{2\pi}{|\theta|} \,.
  \label{debye}
\end{equation}
This strongly suggests that $\theta$ is restricted to zero in the confining phase of the theory, which would mean that the strong CP problem is solved by itself.

A model-independent evaluation of the IR behavior of QCD has remained elusive. This is a multi-scale problem, which involves the passage from the short-distance weakly coupled regime, the lattice, to the long-distance strongly coupled confinement regime. The framework for dealing with physical problems involving different energy scales is the multi-scale renormalization group (RG) flow. Exact RG transformations are very difficult to implement numerically. The gradient flow~\cite{Narayanan:2006rf,Luscher:2010iy} provides a powerful alternative for scale setting, with no need for costly ensemble matching. It can be regarded as a particular, infinitesimal realization of the coarse-graining step of momentum space RG transformations~\cite{Luscher:2013vga,Makino:2018rys,Abe:2018zdc,Carosso:2018bmz} \`a la Wilson~\cite{Wilson:1973jj}, Polchinski~\cite{Polchinski:1983gv} and Wetterich~\cite{Berges:2000ew}, which leaves the long-distance physics unchanged.

In this talk I will review recent work~\cite{Nakamura:2021meh} on the IR behavior of the theory in the presence of the $\theta$ term (\ref{charge}) using the gradient flow, and show that CP is naturally conserved in the confining phase of QCD.

\section{Gradient flow}
\label{sec2}

The gradient flow describes the evolution of fields and physical quantities as a function of flow time $t$. The flow of SU(3) gauge fields is defined by~\cite{Luscher:2010iy}
\begin{equation}
\partial_{\,t}\,B_\mu(t,x) = D_\nu \, G_{\mu\nu}(t,x) \,, \quad G_{\mu\nu} = \partial_\mu\,B_\nu -\partial_\nu\,B_\mu + [B_\mu, B_\nu] \,, \quad D_\mu\, \cdot = \partial_\mu \cdot + \,[B_\mu,\cdot]\,,
\label{gflow}
\end{equation}
where $B_\mu(t,x) = B_\mu^{\,a}(t,x)\,T^a$, and $B_\mu(t=0,x) = A_\mu(x)$ is the original gauge field of QCD. It thus defines a sequence of gauge fields parameterized by $t$. The renormalization scale $\mu$ is set by the flow time, $\mu=1/\sqrt{8t}$ for $t \gg 0$, where $\sqrt{8t}$ is the `smoothing range' over which the gauge field is averaged. The expectation value of the energy density 
\begin{equation}
   E(t,x) = \frac{1}{2}\, \mathrm{Tr}\, G_{\mu\nu}(t,x)\,  G_{\mu\nu}(t,x) = \frac{1}{4}\, G_{\mu\nu}^a(t,x)\,  G_{\mu\nu}^a(t,x) 
  \label{energydensity}
\end{equation}
defines a renormalized coupling 
\begin{equation}
g_{GF}^2(\mu) =  \frac{16 \pi^2}{3}\, t^2 \langle E(t) \rangle\,\big|_{\,t=1/8\mu^2}
\label{gfcoupling}
\end{equation}
at flow time $t$ in the gradient flow (GF) scheme. Varying $\mu$, the coupling satisfies standard RG equations. 
  
We restrict our investigations to the SU(3) Yang-Mills theory. If the strong CP problem is resolved in the Yang-Mills theory, then it is expected that it is also resolved in QCD for massive quarks. We use the plaquette action
\begin{equation}
S_0 = \beta \sum_{x,\,\mu < \nu} \Big( 1 - \frac{1}{3}\, \mathrm{Re}\, \mathrm{Tr}\; U_{\mu\nu}(x)\Big)\,, \quad \beta=\frac{6}{g^2}
\end{equation}
to generate representative ensembles of fundamental gauge fields. For any such gauge field the flow equation (\ref{gflow}) is integrated to the requested flow time $t$. We use a continuum-like version of the energy density $E(t,x)$ obtained from a symmetric (clover-like) definition of the field strength tensor $G_{\mu\nu}(t,x)$~\cite{Luscher:2010iy}. The simulations are done for $\beta=6.0$ on $16^4$, $24^4$ and $32^4$ lattices. The lattice spacing at this value of $\beta$ is $a=0.082(2) \, \mathrm{fm}$, taking $\sqrt{t_0}=0.147(4)\,\mathrm{fm}$ to set the scale~\cite{Bornyakov:2015eaa,Miller:2020evg}, where $t_0$ is implicitly defined by $t_0^2\, \langle E(t_0)\rangle=0.3$. Our current ensembles include $4000$ configurations on the $16^4$ lattice and $5000$ configurations on the $24^4$ and $32^4$ lattices each. The flowed fields are recorded every $\Delta t/a^2 = 1$, $5$ and $10$ on the $16^4$, $24^4$ and $32^4$ lattices, respectively.

\begin{figure}[!b]
  \vspace*{-1.25cm}
  \begin{center}
 \includegraphics[width=9cm]{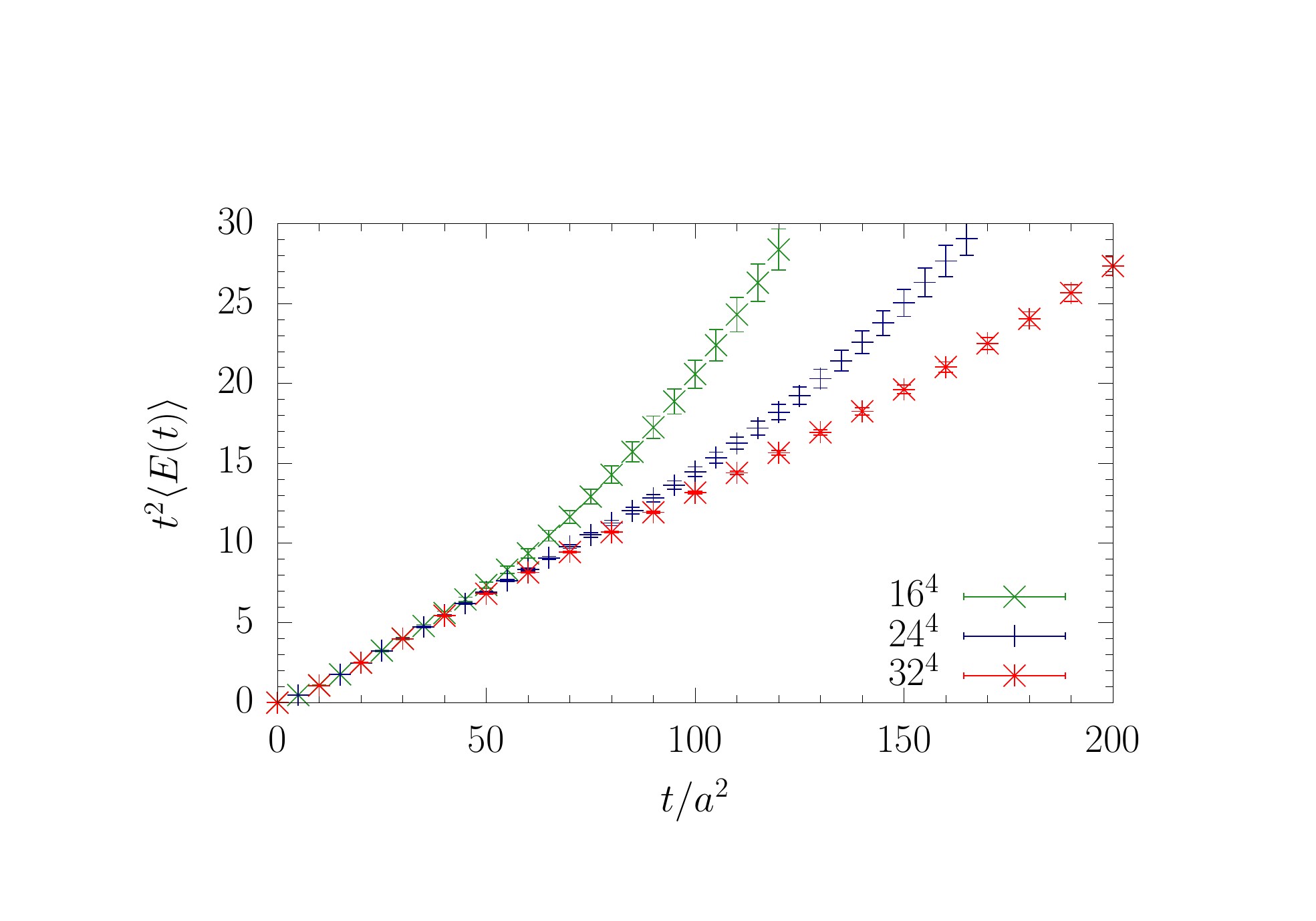}
  \end{center}
  \vspace*{-1.25cm}
\caption{The dimensionless quantity $t^2 \langle E(t)\rangle$ as a function of $t/a^2$ on the $16^4$, $24^4$ and $32^4$ lattice.}
\label{fig3}
\vspace*{-0.25cm}
\end{figure}

In this talk I will consider bulk quantities only, like the energy density $E(t)$, for example. Limits on the flow time are set by the lattice volume. For bulk quantities we expect finite size effects to become noticeable at $\sqrt{8t} \gtrsim L$, where $L$ is the linear extent of the lattice. To check this I plot the dimensionless quantity $t^2 \langle E(t)\rangle$ on the $16^4$, $24^4$ and $32^4$ lattice as a function of $t/a^2$ in Fig.~\ref{fig3}. The data on the $32^4$ lattice fall on a straight line up to $t/a^2 \approx 150$, corresponding to $\sqrt{8t} \approx 32$. On the $24^4$ and $16^4$ lattices finite size effects show up at $t/a^2 \approx 80$ and $50$, respectively, also in rough agreement with $\sqrt{8t} \approx L$.

\begin{figure}[!t]
  \vspace*{-1.25cm}
  \begin{center}
    \includegraphics[width=9cm]{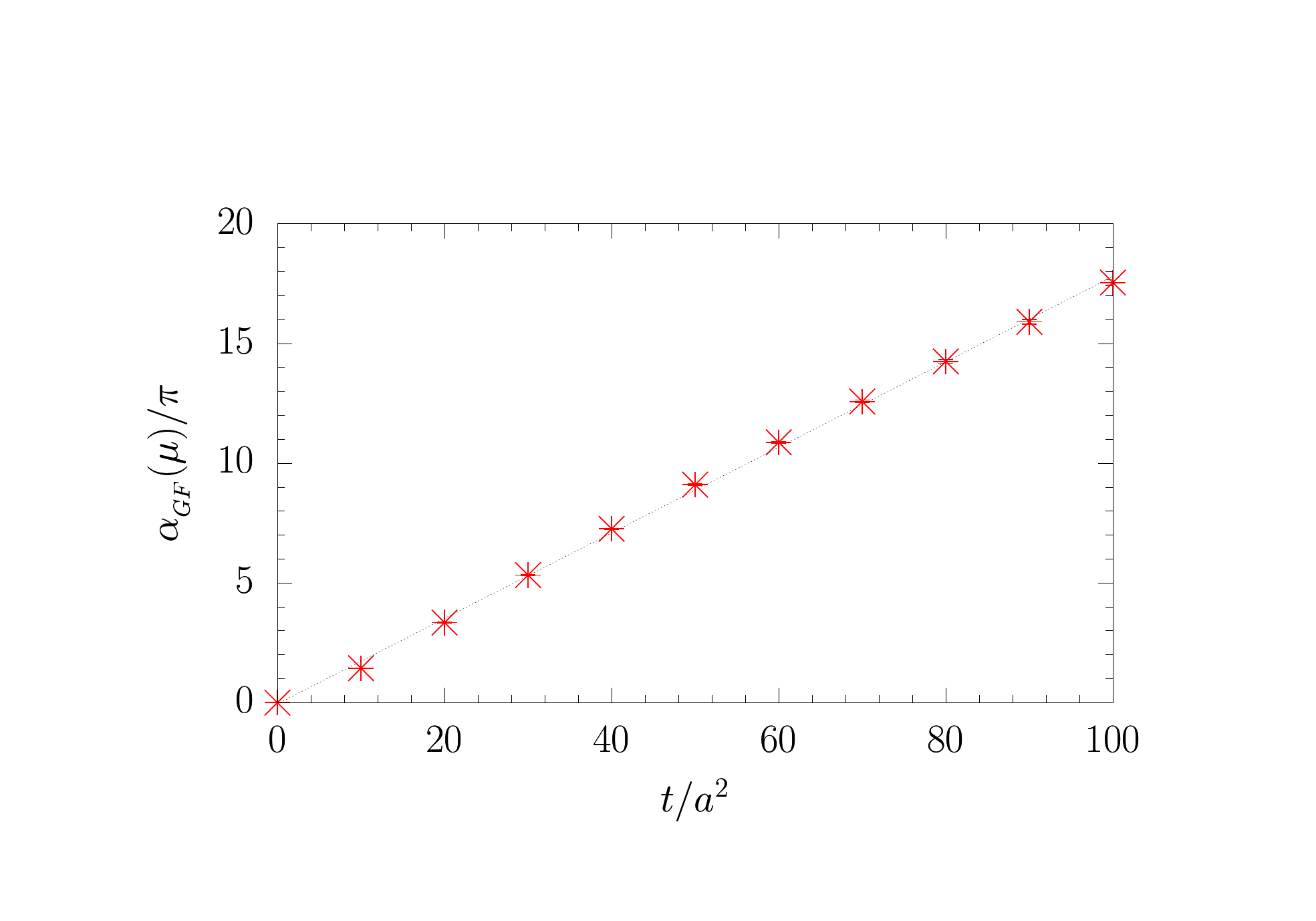}
  \end{center}
  \vspace*{-1.25cm}
\caption{The gradient flow coupling $\alpha_{GF}(\mu)/\pi$ on the $32^4$ lattice as a function of $t/a^2=1/8a^2\mu^2$, together with a linear fit.}
\label{fig4}
\vspace*{-0.25cm}
\end{figure}

\section{Linear confinement}
\label{sec3}

The gradient flow running coupling $\alpha_{GF}(\mu)=g_{GF}^2(\mu)/4\pi$, introduced in (\ref{gfcoupling}), plays a key role in our investigations. In Fig.~\ref{fig4} I show $\alpha_{GF}/\pi$ on the $32^4$ lattice as a function of $t/a^2=1/8\,a^2\mu^2$. In Fig.~\ref{fig3} we have already seen that the linear behavior extends to $\sqrt{8t} \approx L$, corresponding to $\mu \lesssim 100\, \textrm{MeV}$, and it may be assumed that it will extend linearly to even larger values of $t$ as the volume is increased. This gives rise to the beta function
\begin{equation}
    \frac{\partial\, \alpha_{GF}(\mu)}{\partial\, \ln \, \mu}  \equiv \beta_{GF}(\alpha_{GF}) \underset{\mu^2\, \ll\, 1\,\mathrm{GeV}^2}{=} -\, 2 \, \alpha_{GF}(\mu)\,,
    \label{rgGF}
\end{equation}
which has the implicit solution
\begin{equation}
  \frac{\Lambda_{GF}}{\mu}=\exp\left\{-\int_{\alpha_{GF}(\Lambda_{GF})}^{\alpha_{GF}(\mu)} \! d\alpha  \, \frac{1}{\beta_{GF}(\alpha)}\right\}\,, \quad \alpha_{GF}(\mu) \underset{\mu^2\, \ll\, 1\,\mathrm{GeV}^2}{=} \frac{\Lambda_{GF}^2}{\mu^2} \,.
  \label{lambdagf}
\end{equation}
A fit to the lattice data gives $\sqrt{t_0}\,\Lambda_{GF} = 0.475(16)$. It is understood that the beta function connects analytically to the perturbative expression. The linear increase of the running coupling is a universal property of the theory~\cite{Nakamura:2021meh}. In any other scheme $S$
\begin{equation}
  \alpha_{S}(\mu) \underset{\mu^2\, \ll\, 1\,\mathrm{GeV}^2}{=} \frac{\Lambda_{S}^2}{\mu^2}\,.
\end{equation}

To make contact with phenomenology, we need to transform the $\Lambda$ parameter to a matching scheme. Popular schemes are the $V$ (potential) scheme and the $\ols{MS}$ scheme. Knowing $\Lambda_V/\Lambda_{\ols{MS}}=1.600$~\cite{Schroder:1998vy} and $\Lambda_{\ols{MS}}/\Lambda_{GF}=0.534$~\cite{Luscher:2010iy}, we obtain $\sqrt{t_0}\,\Lambda_V=0.406(14)$ and $\sqrt{t_0}\,\Lambda_{\ols{MS}}=0.217(7)$. The latter number is in excellent agreement with the result of a recent dedicated lattice calculation~\cite{DallaBrida:2019wur}, $\sqrt{t_0}\, \Lambda_{\ols{MS}} = 0.220(3)$. The linear increase of $\alpha_V(\mu)=\Lambda_V^2/\mu^2$ with $1/\mu^2$ is commonly dubbed infrared slavery. It effectively describes many low-energy phenomena of the theory. So, for example, the static quark-antiquark potential, which can be described by the exchange of a single dressed gluon, $V(q)= -\frac{4}{3}\,\alpha_V(q)/q^2$. Upon performing the Fourier transformation of $V(q)$ to configuration space, we have
\begin{equation}
  V(r) = -\frac{1}{(2\pi)^3} \int d^3\mathbf{q} \; e^{i\,\mathbf{q r}} \; \frac{4}{3}\, \frac{\alpha_V(q)}{\mathbf{q}^2 + i 0}\; \underset{r\, \gg\, 1/\Lambda_V}{=} \,\sigma \, r\,,
  \label{pot}
\end{equation}
where $\sigma$, the string tension, is given by $\sigma = \frac{2}{3}\, \Lambda_V^2$. The result is $\sqrt{t_0\,\sigma}=0.331(11)$. Converted to physical units, we obtain $\sqrt{\sigma} = 445(19)\,\mathrm{MeV}$, which is exactly what one expects from Regge phenomenology. Both results,  $\Lambda_{\ols{MS}}$ and $\sigma$, are to be considered a success for our approach.

\section{Topological freezing}
\label{sec4}

It is well known that in the continuum limit the path integral splits into quantum mechanically disconnected sectors of topological charge $Q$. It so happens that the ensemble of gauge fields freezes into isolated topological sectors with increasing flow time. A similar behavior has been found long time ago by `cooling' lattice gauge field configurations~\cite{Ilgenfritz:1985dz,Bonati:2014tqa}. This is expected to occur at ever smaller flow times as the lattice spacing is reduced~\cite{Luscher:2010iy}. 

\begin{figure}[!h]
  \vspace*{-1.20cm}
  \begin{center}
 \hspace*{-0.4cm}   \includegraphics[width=9.2cm]{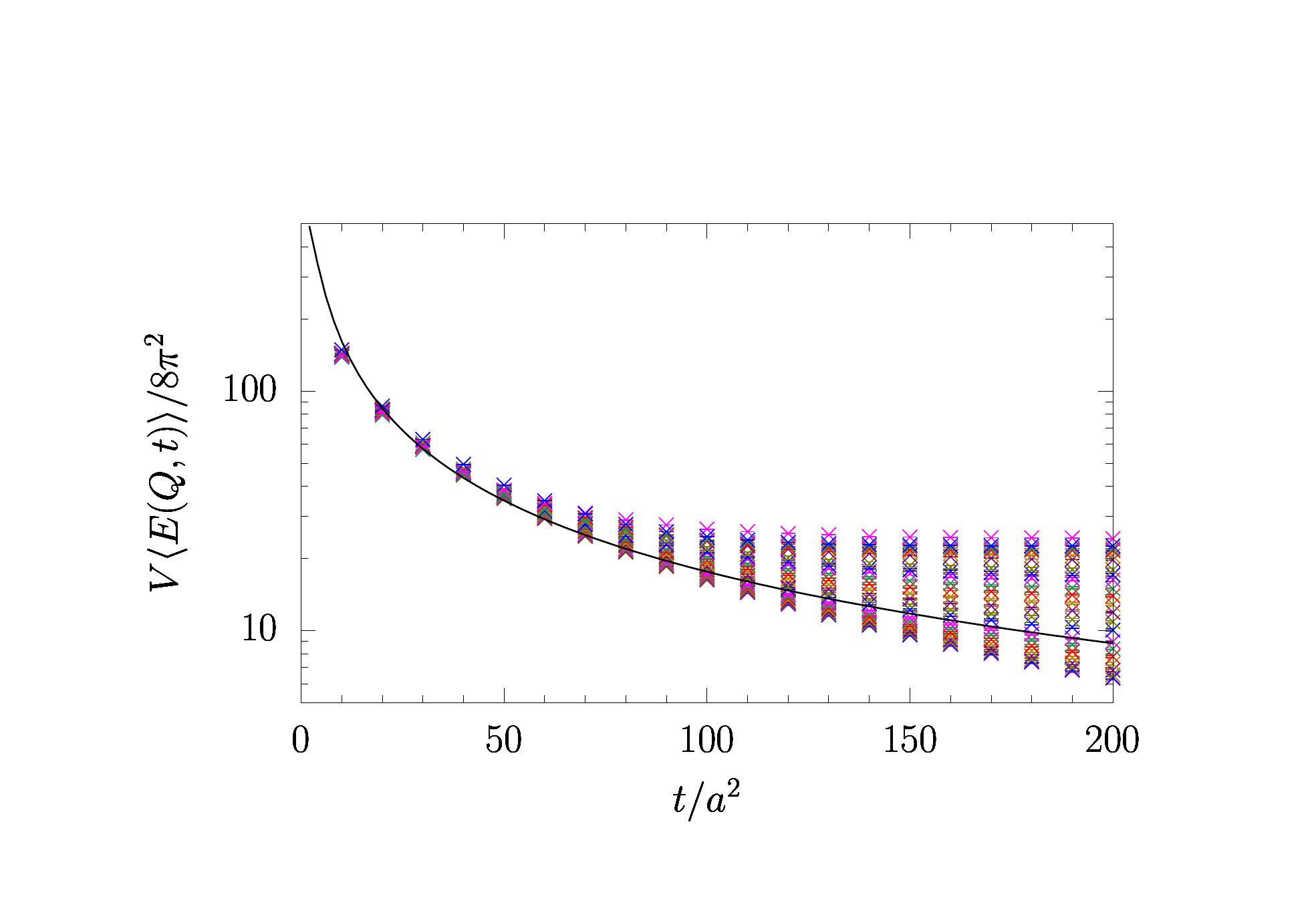}
  \end{center}
  \vspace*{-1.25cm}
\caption{The action $V\langle E(Q,t)\rangle/8\pi^2$ as a function of $t/a^2$ on the $32^4$ lattice for charges ranging from $Q=0$ (bottom) to $|Q|=22$ (top). The solid line represents the ensemble average. No transition from one sector to the other is observed.}
\label{fig5}
\vspace*{-0.25cm}
\end{figure}

We distinguish the topological sectors by the affix $Q$, where $Q$ is defined by the lattice version of $Q = (1/32\pi^2)\, \epsilon_{\mu\nu\rho\sigma} \int d^4x\, \textrm{Tr}\left[G_{\mu\nu} G_{\rho\sigma}\right]$. A prominent example is the energy density $\langle E(Q,t)\rangle$. In Fig.~\ref{fig5} I plot the average `action' $V\langle E(Q,t) \rangle/8\pi^2$, normalized to one for a single classical instanton, on the $32^4$ lattice as a function of $t/a^2$. While $V\langle E(Q,t)\rangle/8\pi^2$ develops a plateau for borderline charges at large flow time, the ensemble average, that is the statistical average across all topological sectors, vanishes like $1/t$. Unlike `cooling', the topological sectors are remarkably stable, even for borderline charges.

\begin{figure}[!b]
  \vspace*{-0.75cm}
  \begin{center}
\hspace*{0.6cm}\includegraphics[width=9cm]{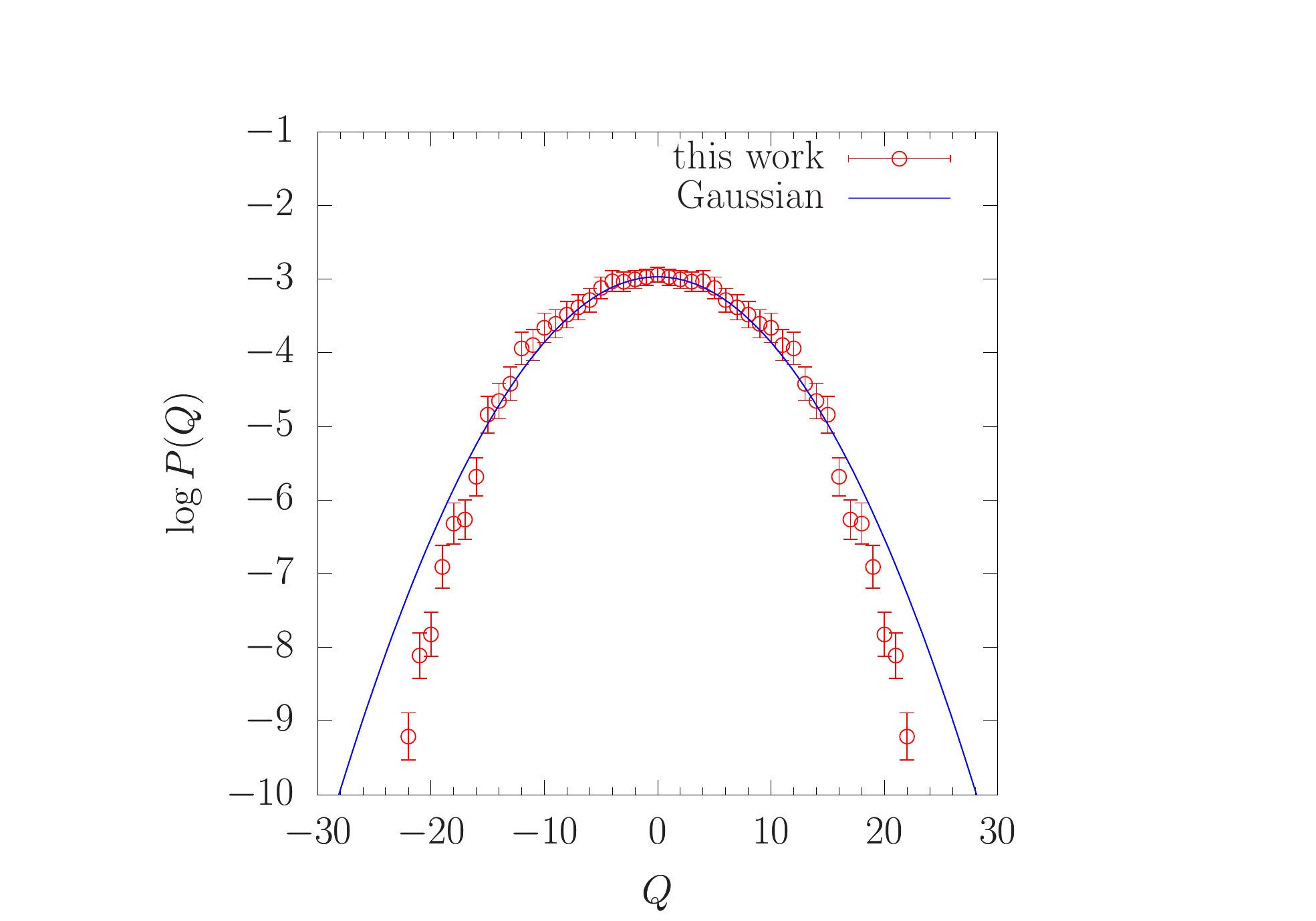}
  \end{center}
  \vspace*{-0.5cm}
\caption{The logarithm of the topological charge distribution $P(Q)$ on the $32^4$ lattice, compared to a Gaussian distribution.}
\label{fig6}
\vspace*{-0.5cm}
\end{figure}

One might think that the gradient flow ends up in a semi-classical ensemble of  noninteracting instantons. This is, however, not the case. Under the gradient flow $\partial_t Q=0$~\cite{Nakamura:2021meh}. It then follows that the probability distribution for topological charge,
\begin{equation}
  P(Q) = \frac{\int_Q\mathcal{D} U_\mu\, e^{-S_0}}{\int\mathcal{D} U_\mu\, e^{-S_0}}\,,
  \label{PQ}
\end{equation}
is independent of the flow time $t$, once the ensemble has settled into disconnected topological sectors. Furthermore, $P(Q)$ deviates significantly from the distribution of a dilute instanton gas, not to mention a Gaussian distribution. In Fig.~\ref{fig6} I compare the logarithm of $P(Q)$ with a Gaussian distribution on the $32^4$ lattice. The distribution turns out to be much narrower than Gaussian, which indicates repulsion of instantons~\cite{Shuryak:1995pv}. From $P(Q)$ we obtain the topological susceptibility, $\chi_t=\sum_Q Q^2 P(Q)/V$. On the $32^4$ lattice we find $\sqrt{t_0}\:\chi_t^{1/4}= 0.162(3)$, which agrees precisely with the value $\sqrt{t_0}\:\chi_t^{1/4}=0.161(4)$ reported in~\cite{Ce:2015qha}.

\begin{figure}[!b]
  \vspace*{-0.75cm}
  \begin{center}
\hspace*{1.75cm}\includegraphics[width=9.5cm]{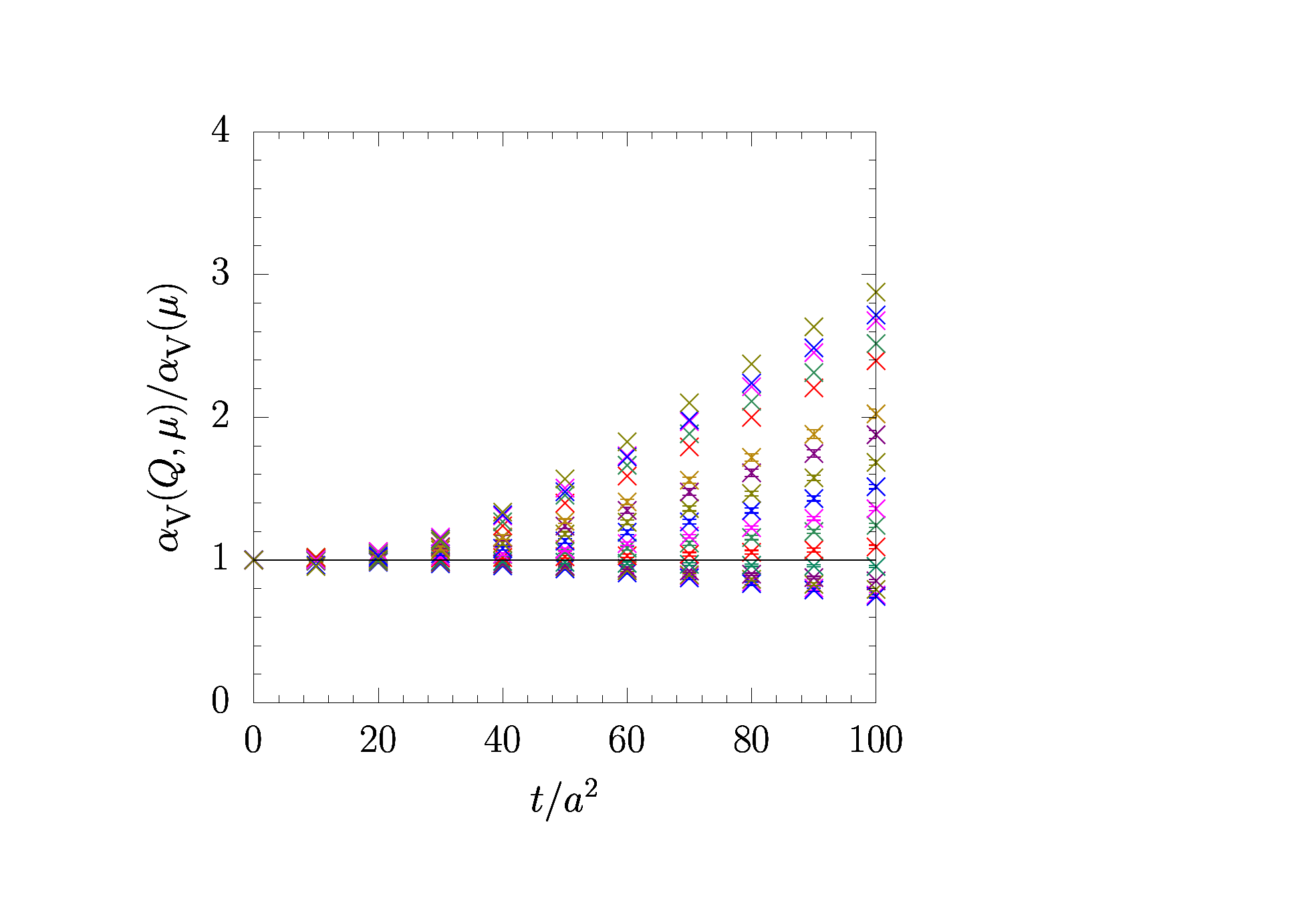}
  \end{center}
  \vspace*{-1cm}
\caption{The ratio $\alpha_V(Q,\mu)/\alpha_V(\mu)$ as a function of $t/a^2$ on the $24^4$ lattice for charges ranging from $Q=0$ (bottom) to $|Q|=16$. No errorbars are shown for marginal values of $Q$ because of limited statistics.}
\label{fig7}
\vspace*{-0.25cm}
\end{figure}

\section{Effect of {\bf $\theta$} term}
\label{sec5}

If the color fields are screened in the $\theta$ vacuum, this is expected to show in the $\theta$-dependence of the running coupling in the first place. From $\langle E(Q,t)\rangle$ we derive $\alpha_V(Q,\mu)$ depending on $Q$. In Fig.~\ref{fig7} I show $\alpha_V(Q,\mu)$ divided by the ensemble average $\alpha_V(\mu)$ on the $24^4$ lattice, where we have the best data. Already at relatively small flow times $\alpha_V(Q,\mu)$ begins to fan out according to $Q$. The transformation to $\theta$ is achieved by the discrete Fourier transform
\begin{equation}
    \alpha_V(\theta,\mu) = \frac{1}{Z(\theta)} \sum_Q e^{\,i\,\theta\, Q}\, P(Q)\; \alpha_V(Q,\mu)\,,\quad Z(\theta) = \sum_Q e^{\,i\,\theta\, Q}\, P(Q)\,,    \label{fouriert}
\end{equation}
weighted by the probability distribution $P(Q)$ in (\ref{PQ}). Here $\theta$ is the {\em bare} vacuum angle, labelling superselection sectors. It is the parameter that appears in the lattice action and determines the phases of the theory. Limits are set by the precision of $P(Q)$ and $\alpha_V(Q,\mu)$. The charge distribution $P(Q)$ is determined by the real part of the action, $S_0$, which increases with $|Q|$ and, thus, suppresses configurations with a large number of (anti-)instantons. That makes it increasingly difficult to determine $P(Q)$ precisely for large values of $|Q|$. For our main conclusions we need to know $\alpha_V(\theta,\mu)$ for small values of $|\theta|$ only, which is rather insensitive to fluctuations at large values of $|Q|$. 

\begin{figure}[!t]
  \begin{center}
\hspace*{-0.5cm}\includegraphics[width=6cm]{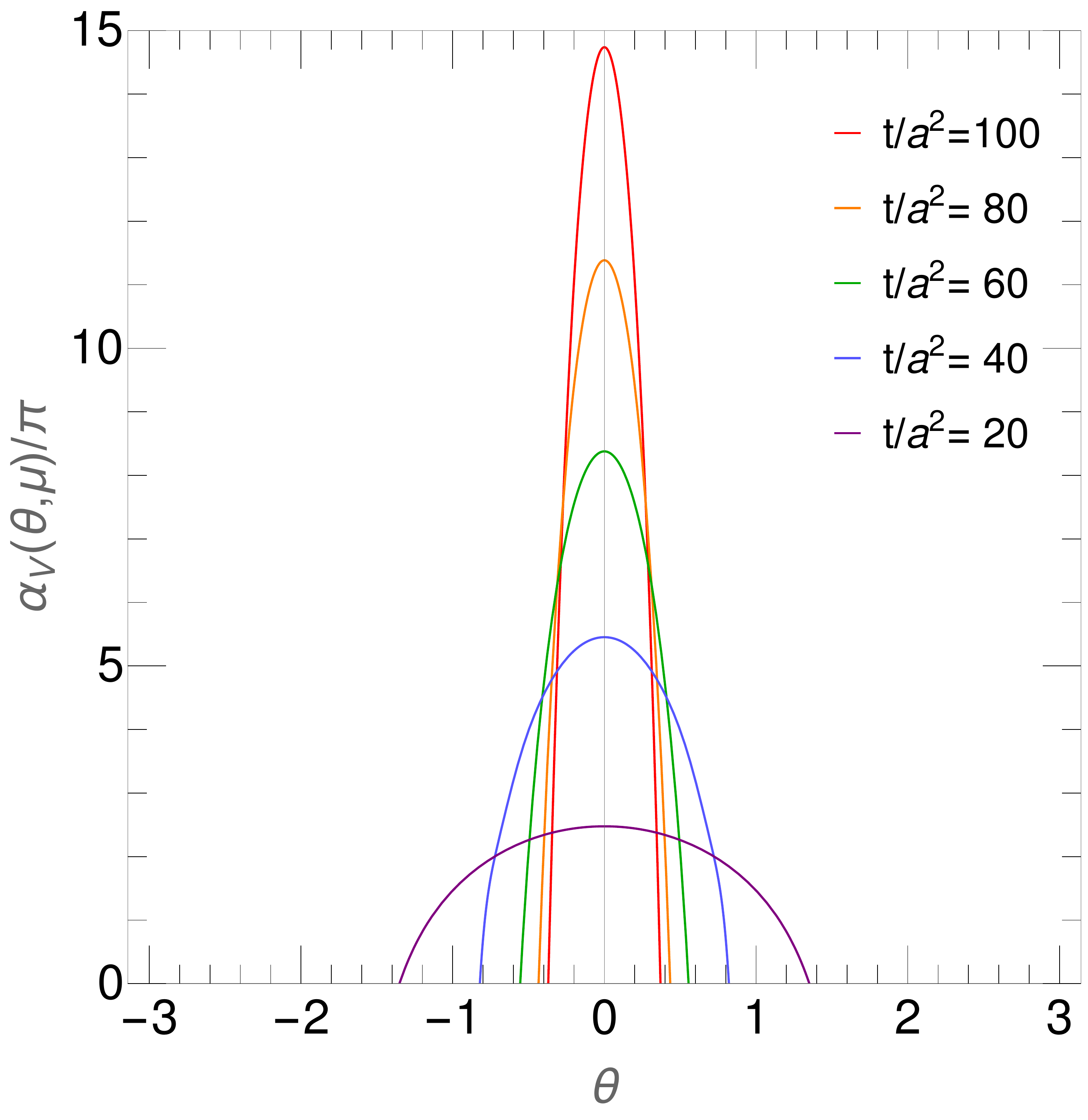}
  \end{center}
  \vspace*{-0.5cm}
\caption{The running coupling $\alpha_V(\theta,\mu)$ as a function of $\theta$ on the $24^4$ lattice for flow times from $t/a^2=20$ (bottom) to $100$ (top).}
\label{fig8}
\vspace*{-0.25cm}
\end{figure}

Figure~\ref{fig8} shows $\alpha_V(\theta,\mu)$. Similar results are obtained on the $16^4$ and $32^4$ lattices~\cite{Nakamura:2021meh}. The calculations of $\alpha_V(\theta,\mu)$ are rather robust. The determining factor is that $\alpha_V(Q,\mu)$ is a monotonically increasing function of $|Q|$, which makes the numerator of (\ref{fouriert}) fall off much faster than the denominator, $Z(\theta)$, while the charge distribution $P(Q)$ largely cancels out. With our current statistics we are not able to compute $\alpha_V(\theta,\mu)$ with confidence for $t/a^2 \lesssim 10$ on the $24^4$ lattice. But there is no doubt that it will continue to flatten. The main hindrance is that fluctuations of $\alpha_V(Q,\mu)$ with respect to $Q$ can be significant on larger volumes before the ensemble has settled into truly disconnected topological sectors. The situation is expected to improve for smaller lattice spacings $a$.  

\begin{figure}[!t]
  \vspace*{-1.25cm}
  \begin{center}
\includegraphics[width=9.5cm]{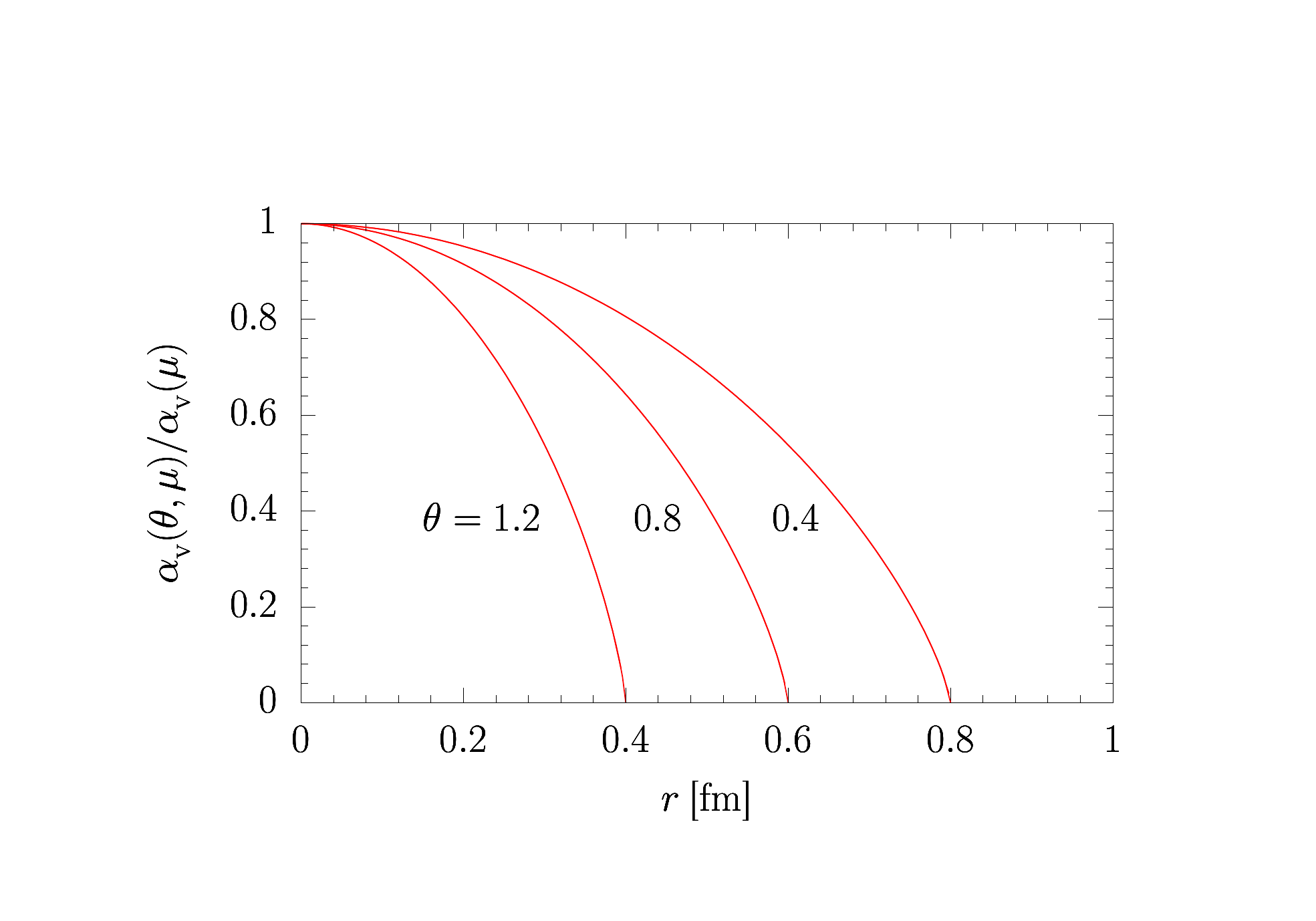}
  \end{center}
  \vspace*{-0.75cm}
\caption{The running coupling $\alpha_V(\theta,\mu)$ divided by the ensemble average $\alpha_V(\mu) \equiv \alpha_V(0,\mu)$ as a function of distance for several values of $\theta$.}
\label{fig9}
\vspace*{-0.25cm}
\end{figure}

For any fixed value $|\theta| > 0$ the running coupling $\alpha_V(\theta,\mu)$ appears to be bounded by a finite value, however small $|\theta|$ is, which thwarts linear confinement. When viewed from a fixed value of flow time, we find that $\alpha_V(\theta,\mu)$ drops to zero as $|\theta|$ is increased. This happens the faster, the larger $t$ is. The scale parameter $\mu$ corresponds to the distance $r = \exp\{-\gamma_E\}/\sqrt{2}\, \mu \approx 0.40/\mu$ at which the color charge is probed, which derives from converting the potential $V(q)$ to $V(r)$ in coordinate space~\cite{Necco:2003jf}. The curves in Fig.~\ref{fig8} thus acquire a concrete physical meaning. For example, at $|\theta| = 0.8$ the color charge is totally screened at distances $r \gtrsim 0.6\;\textrm{[fm]}$, while at $|\theta| = 0.4$ it is screened for $r \gtrsim 0.8\;\textrm{[fm]}$. This means that for any fixed value $|\theta|>0$ quarks and gluons can be separated, perhaps with increasing cost of energy. Analytically, $\alpha_V(\theta,\mu)$ in Fig.~\ref{fig8} can be expressed by~\cite{Nakamura:2019ind}
\begin{equation}
  \alpha_V(\theta,\mu) = \alpha_V(\mu) \left[1-\alpha_V(\mu)\, (D/\lambda)\, \theta^2\right]^\lambda 
  \label{alphaex}
\end{equation}
within the errors. On the $24^4$ lattice $D \approx 0.10$ and $\lambda \approx 0.75$. The charge is totally screened when the right-hand side vanishes. Figure~\ref{fig9} shows how the running coupling $\alpha_V(\theta,\mu)$ is screened with increasing distance for several values of $\theta$. We define the screening radius, $\lambda_S$, at which the running coupling has dropped by a factor $1/e$. From (\ref{alphaex}) we find
\begin{equation}
  \lambda_S \approx \frac{0.31}{|\theta|} \;\textrm{[fm]} \,.
  \end{equation}
This result agrees with the Debye length (\ref{debye}) of the superconductor model for $\sqrt{E_F/\rho} \approx 2\,\textrm{fm}$. The situation is very similar to the finite temperature phase transition. While confinement is lost for $T > T_c$, the screening length~\cite{Kaczmarek:2004gv} is consistently described by $\lambda_S \,\propto\,1/(T-T_c)$. 

\section{Renormalization of the {\bf $\theta$} angle}
\label{sec6}

\begin{figure}[!b]
  \begin{center}
\includegraphics[width=7cm]{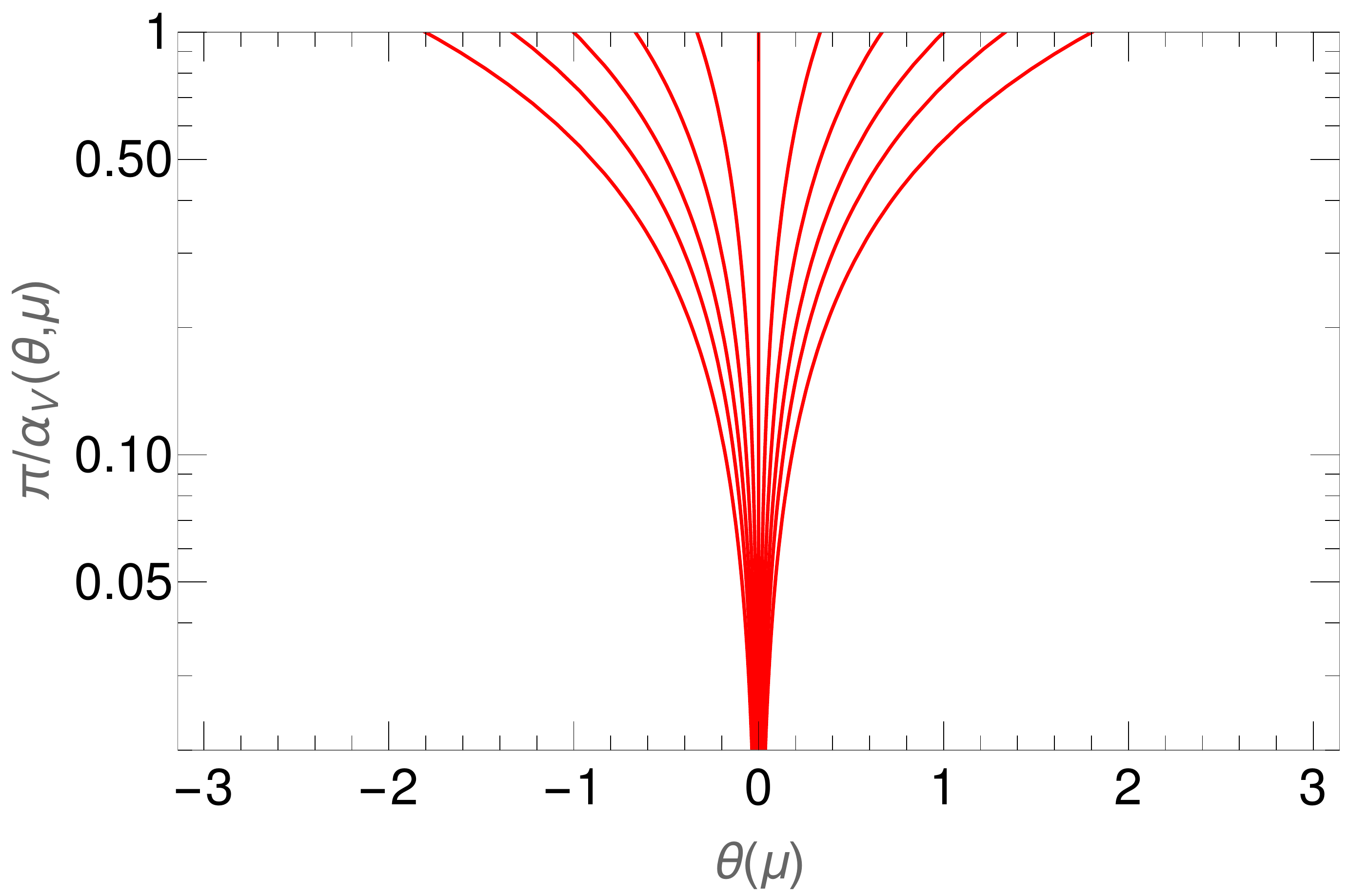}
  \end{center}
  \vspace*{-0.6cm}
\caption{The RG flow of $\alpha_V(\theta,\mu)$ and $\theta(\mu)$ in the $(\theta, \pi/\alpha_V)$ plane for different initial values of $\theta$, with $t$ increasing from top to bottom.}
\label{fig10}
\vspace*{-0.25cm}
\end{figure}

From Fig.~\ref{fig8} it is evident that {\em within} QCD, $\theta$ can only assume values inside the envelopes of the curves. The conclusion is that $\theta$ gets renormalized in the IR, together with $\alpha_V$. From (\ref{alphaex}) we can derive RG equations, which determine the flow of $\alpha_V$ and $\theta$ as a function of the scale parameter $\mu$. This leads to a set of partial differential equations. For $\alpha_V \theta^{\,4} \ll 1$ the equations simplify to
\begin{equation}
  \frac{\partial\, [1/\alpha_V(\theta,\mu)]}{\partial \ln t} \simeq - \frac{1}{\alpha_V(\theta,\mu)} + D\, \theta(\mu)^2\,, \quad  \frac{\partial\, \theta(\mu)}{\partial \ln t} \simeq - \frac{1}{2}\, \theta(\mu)\,,
  \label{RGeqs}
\end{equation}
which applies to the major part of Fig.~\ref{fig10}. In (\ref{RGeqs}) $\theta(\mu)$ denotes the {\em renormalized} $\theta$ parameter, which can be thought of as the coupling that enters the effective Lagrangian averaged over virtualities larger than $\mu$. In Fig.~\ref{fig10} I show the RG flow of $\alpha_V(\theta,\mu)$ and $\theta(\mu)$ in the $(\theta,\pi/\alpha_V)$ plane for different initial values of $\theta$, with $t$ increasing from top to bottom. Assuming that there is no phase transition in the IR limit, we conclude that both $\theta(\mu)$ and $\pi/\alpha_V(\theta,\mu)$ flow to zero in the limit $\mu \rightarrow 0$, very likely to an IR fixed point. Bar of any loop corrections, the properties of the theory can be directly read off from the fixed point couplings, to the end that confinement implies strong CP invariance. At the upper end of the curves, that is in the perturbative regime, CP is trivially conserved. This suggest that CP is conserved at all scales $\mu$ {\em within} QCD. Actually, to be true, the vacuum expectation value $\langle G_{\mu\nu}^a\tilde{G}_{\mu\nu}^a\rangle_\theta$ must be zero everywhere for $|\theta|>0$~\cite{Shifman:1979if}. In~\cite{Nakamura:2021meh} we have argued that this is indeed the case.

The predictions of~\cite{Pruisken:2000my,Levine:1983vg,Knizhnik:1984kn}, stating that the $\theta$ parameter renormalizes to zero [mod $2\pi$] in the IR limit, has thus proven true. Remarkably, the RG equations derived from the effective Lagrangian, based on the dilute instanton gas~\cite{Callan:1977gz}, are largely equivalent to ours (\ref{RGeqs}), considering the rather limited size of instantons~\cite{Shuryak:1995pv}. Last not least, it has been shown analytically~\cite{Reuter:1996be} that $\theta(0) = 0$ when $\alpha_S(0) = \infty$, based on an exact RG evolution equation~\cite{Reuter:1993kw}. This coincides with our result.

\vspace*{-0.1cm}

\section{Lesson for axions}
\label{sec7}

The Peccei-Quinn proposal for the solution of the strong CP problem replaces the static $\theta$ angle with a dynamical CP-conserving field, the axion. While the axion is certainly not needed for the solution, the question is whether the model is consistent with our results, and QCD in general. In this model the action is augmented by the axion interaction
\begin{equation}
  S_{\textrm{Axion}} = \int \! d^4x \, \left[\frac{1}{2} \big(\partial_\mu \phi_a(x)\big)^2 + i \left(\theta + \frac{\phi_a(x)}{f_a}\right) \, q(x) \right]\,, \quad Q = \int d^4x\, q(x) \,,
\end{equation}
where interactions of derivatives of $\phi_a$ with matter fields have been suppressed. Formally, the path integral
\begin{equation}
  \int \mathcal{D}\phi_a\,\mathcal{D}A_\mu\, \exp\{-S_0-S_{\textrm{Axion}}\}
\end{equation}
is invariant under the shift of integration variables $\phi_a \rightarrow \phi_a+\delta$, called shift symmetry. This is the crux of the whole idea. It is invariant as long as the integration extends from $\bar{\phi}_a = -\infty$ to $\bar{\phi}_a = +\infty$, with $\bar{\phi}_a = (1/V) \int d^4x\, \phi_a(x)$. However, this is not the case. If the gauge fields are integrated first, $\bar{\phi}_a$ is restricted to an increasingly narrow region around $\bar{\phi}_a=0$. If, on the other hand, the axion field is integrated first, the topological charge is fixed at $Q=0$~\cite{GS}, which thwarts chiral symmetry breaking, the pseudoscalar meson spectrum, and probably is inconsistent with confinement. 

\vspace*{-0.1cm}

\section{Conclusions}
\label{sec8}

There is no space left to discuss errors. For an error estimate, and an assessment of finite volume corrections, I refer the interested reader to our original paper~\cite{Nakamura:2021meh}. 

The gradient flow proved a powerful tool for addressing the nonperturbative, long-distance properties of QCD. It showed its potential for extracting low-energy quantities of the theory, highlighted by the topological susceptibility $\chi_t$, the lambda parameter $\Lambda_{\ols{MS}}$ and the string tension $\sigma$. Under the gradient flow the path integral splits into quantum mechanically disconnected topological sectors of charge $Q$. An important observation is that the charge density $P(Q)$, that is the probability of finding a vacuum configuration of charge $Q$, is independent of the flow time $t$. Thus, all quantities that derive from the free energy, $F(\theta) = (1/V)\log{Z(\theta)}$, like cumulants of $Q$ and the $\eta^\prime$ mass and self-coupling~\cite{Kronfeld:1988vc}, are independent of $t$ as well, and the vacuum does not collapse to a dilute gas of instantons.

So far we have concentrated on bulk quantities, specifically the energy density. The energy density defines a renormalized coupling in the gradient flow scheme. For phenomenological reasons we converted the coupling to the $V$ scheme, $\alpha_V$. The renormalized coupling reflects the color charge of a quark or gluon probed at a given distance. In ordinary QCD, at $\theta=0$, the charge attracts other charges of the same color from the vacuum, an effect called anti-screening. At this point $\alpha_V$ was found to increase quadratically with distance, in accord with infrared slavery. For $|\theta| > 0$ the picture turns and the opposite effect is found. As $|\theta|$ is increased, the color charge is screened, to the extent that the renormalized coupling $\alpha_V(\theta,\mu)$ vanishes at distances inversely proportional to $|\theta|$. To maintain confinement, the $\theta$ parameter needs to be renormalized as well, $\theta = \theta(\mu)$. The screening mechanism can be understood from the dual superconductor model of confinement. In this model anti-screening arises from monopole condensation. By contrast, for nonvanishing values of $\theta$ the monopoles acquire a color(-electric) charge $\propto \,\theta$, driving the vacuum gradually into a Coulomb or Higgs phase, which repeals superconductivity, and screens the color charge instead. The running of the renormalized coupling $\alpha_V(\theta,\mu)$ and the vacuum angle $\theta(\mu)$ can be quantified by RG equations. In the IR limit, $\mu=0$, they have the solution $1/\alpha_V = \theta = 0$, possibly marking an infrared fixed point. This suggests that CP is conserved under the strong interactions.

\vspace*{-0.1cm}
\section*{Acknowlegment}

I like to thank the organizers of the `XV Quark Confinement and the Hadron Spectrum Conference', Nora Brambilla and Alexander Rothkopf, for inviting me to Stavanger. This work would not have been possible without the contributions of Yoshifumi Nakamura. The numerical calculations have largely been carried out on the HOKUSAI at RIKEN and the Xeon cluster at RIKEN R-CCS using BQCD~\cite{Nakamura:2010qh,Haar:2017ubh}.

%
\end{document}